\documentstyle[12pt,titlepage]{article}
\newcommand{\be}{\begin{equation}}
\newcommand{\ee}{\end{equation}}
\newcommand{\br}{\begin{eqnarray}}
\newcommand{\er}{\end{eqnarray}}
\newcommand{\ds}{\displaystyle}
\newcommand{\half}{\frac{1}{2}}
\def\a{{\alpha}}
\def\b{{\beta}}
\def\g{{\gamma }}
\def\G{{\Gamma}}

\def\s{{\sigma}}
\def\S{{\Sigma}}

\def\m{{\mu}}
\def\n{{\nu}}

\def\dirac#1{\setbox0=\hbox{$#1$}\rlap{\hbox to \wd0{$\hss\mkern1mu/\hss$}}
\box0 }

\begin{document}

\begin{titlepage}
\begin{center}

{\bf Asymptotic Freedom and Infrared slavery in PT-symmetric\\
Quantum Electrodynamics ($\ast $)}\\
\vspace{.2in} \vfill

R.Acharya ($\ast $)\\
Physics Department, Arizona State University, Tempe, AZ 85287\\
\vspace{.1in}
and\\
\vspace{.1in}
P. Narayana Swamy ($\dagger$ )\\
Physics Department, Southern Illinois University, Edwardsville, IL
62026\\
\end{center}
\vspace{.1in}
\begin{center}
{\bf Abstract}
\end{center}

We establish that there is no finite PT-symmetric Quantum
Electrodynamics (QED) and as a consequence the Callan-Symanzik
function $\beta (\alpha) < 0$, for all $\alpha$ greater than zero:
PT-symmetric QED exhibits both asymptotic freedom and infrared
slavery.

\vfill

 \vspace{.7in}

($\dagger$ ): Professor Emeritus, e-mail address: pswamy@siue.edu\\
($\ast $): Professor Emeritus, e-mail address:
Raghunath.acharya@asu.edu\

\noindent June 2005

 \vspace{.1in}

($\ast $) To the memory of Ken Johnson: his masterful
presentations at Harvard,
 Bangalore and Dublin had been a rewarding experience for both of us.

 \vfil

\end{titlepage}

\large
 \noindent \textbf{{1. Introduction}}
\normalsize

In an earlier publication, hereafter referred to as I
[\ref{RAPNS1}], we have demonstrated the absence of a finite
eigenvalue in QED with massless electron, thereby confirming the
result established by Johnson, Baker, Callan et al
[\ref{JBW}-\ref{BakerJohnson}], in the 1970's. The conclusion
arrived at in I was based on a non-perturbative ``gauge technique"
pioneered by Salam [\ref{Delburgo}] and by Ball and Zachariasen
[\ref{Balletal}].

In this note we shall simply transcribe the conclusions of I, to
arrive at an identical conclusion in the PT-symmetric QED,
formulated by Bender and co-workers [\ref{Bender}]. Bender's
PT-QED is obtained by replacing the sign of the fine structure
constant,  $\alpha = e^2/4 \pi$ by $- \alpha$. This theory remains
PT-symmetric and, more importantly, is asymptotically free (AF) !
Our concern in this note is to address the all important question,
namely, is PT-QED also a confining theory? As a first step in
tackling this difficult problem, we shall establish that PT-QED
exhibits infrared (IR) slavery, i.e., the running coupling
$\alpha(q^2)$ increases without bound in the infrared limit, $q^2
\rightarrow 0$. It is now standard lore i.e., the conjecture of
Weinberg [\ref{WeinBook}], Fritzsch, Gell-Mann and Leutwyler and
Gross-Wilczek, then asserts that ``infrared slavery is responsible
for confinement". In the words of Weinberg: ``\emph{the decrease
of the coupling constant at high energy or short distance of
course implies an increase at low energy or long distance}".

Our analysis which closely follows our earlier work in I and the
conclusion stated above relies on establishing the absence of a
finite eigenvalue of $\alpha$ i.e., the PT-QED which is formally
obtained by switching the sign of $\alpha$ does not possess a
non-trivial IR fixed point and consequently does not exhibit a
coulomb phase ! In other words, the Callan-Symanzik function of
PT-QED, $\beta(\alpha)$ as a function of $\alpha$ remains negative
for all values of $\alpha$, for $\alpha > 0$ and never turns over.
This conclusion then reinforces the deep observation of Witten et
al [\ref{Witten}], ``in a conformal field theory, any interacting
field strength must couple both to electrons and to monopoles. In
particular, QED without elementary monopoles cannot have a
non-trivial fixed point". Witten's remark applies equally well to
PT-QED.

Since our entire analysis of PT-QED parallels our earlier
exhaustive discussion of QED in I, line by line, we shall
highlight the salient steps of I (in order to keep the discussion
self-contained), with the important clarification of the role of
the QED transverse vector vertex: in I, we had not explicitly
spelled out the precise forms of the transverse vertex; we do so
here. Throughout our discussion, we will set the ``photon"
propagator to its free form. This is justified since, here,
$\beta(\alpha)= \gamma_A (\alpha)$, where $\gamma_A$ is the
anomalous dimension of the ``photon" and therefore $\beta(\alpha)
= 0$ at the fixed point, hence the ``photon" propagator is bare.
The other important point is the observation (see Bernstein,
[\ref{Bernstein}]) that PT invariance will suffice to yield the
standard form of the electron propagator, as discussed in Sec.2
below.

\vspace{.2in}
 \large
 \noindent \textbf{{2. Dyson-Schwinger equation at  QED
fixed point}}

\vspace{.2in} \normalsize

 We begin by reviewing the salient points of our earlier work, I.
 Let us start with the standard finite QED and
introduce the PT symmetric extension by replacing  $\alpha = e^2/4
\pi$ by $- \alpha$. Indeed the correct starting point for a PT
symmetric QED is the form of the renormalized electron propagator
whose general invariant form
 \be S^{-1}(p) = \dirac p \, A(p^2) + \S(p^2),
\label{Eq2.1} \ee follows from Lorentz invariance and parity
invariance. As a matter of fact PT invariance will suffice to
establish the above invariant form, as has been emphasized by
Bernstein [\ref{Bernstein}]. We shall review the theory of finite
QED in order to provide the appropriate notations even as it
necessitates repeating some material from ref.(\ref{RAPNS1}). In
the finite QED theory, we set $\S(p^2)\equiv 0$, for any $p$ and
chiral symmetry remains an exact symmetry. The proper renormalized
vertex function in QED satisfies the Ward-Takahashi identity \be
(p-p')^{\m}{\G}_{\m}(p,p')=S^{-1}(p)-S^{-1}(p'). \label{Eq2.2} \ee
We would like to refer the reader to our earlier work
[\ref{ourNCpaper}] where we have reviewed the well-known
consequences of the Gell-Mann-Low eigenvalue equation $\psi(x)=0$,
which may or may not have a non-trivial zero at $x_0=\a_0$ at the
position of the bare fine structure constant of QED,
$\a_0=Z_3^{-1}\a$. The important premise of the finite theory of
QED is that the position of the zero, $x=x_0$ can be determined by
working with QED with zero physical mass [\ref{JBW}]. This is
predicated upon the application of Weinberg's theorem
[\ref{Weinberg}], which ensures that terms vanishing
asymptotically in each order of perturbation theory in the massive
case do not sum to dominate over the asymptotic parts.

It can be easily shown that at the Gell-Mann-Low fixed point with
$m=0$ in finite QED, the full, exact, remormalized electron
propagator has the simple scaling form [\ref{AdlerBardeen}] \be
S^{-1}(p) = \dirac p A(p^2)= \dirac p \left (\frac{p^2}{\m^2}
\right)^{\g} \label{Eq2.3} \ee
 where $\g(\a)$ is the anomalous
dimension of the electron in the massless theory given in the
Landau gauge by \be \g=\m \frac{\partial}{\partial \m} \ln Z_2=
O(\a^2)+ \cdots. \label{Eq2.7} \ee and $\m$ is the subtraction
point. This can be established as follows. Starting with the
Callan-Symanzik renormalization group equation [\ref{Callan}], if
we specialize to the Landau gauge,  set $m=0$ (massless electron)
and $\b(\a)=0$ at the fixed point, then we have the equation
satisfied by the two-point function,  essentially the inverse
electron propagator \be \left (\m \frac{\partial}{\partial \m} +
2\g \right )  \G^{(2)}(p, \a, 0, \m,0) = 0 \, . \label{Eq2.4}
 \ee
 The solution for the two-point function can be
expressed as \be A(p^2) = \left (\frac{p^2}{\m^2} \right)^{\g}.
\label{Eq2.5} \ee which   is customarily expressed in terms of
Euclidean momenta.

This can be confirmed by examining the trace anomaly in QED
[\ref{AdlerCollinsDuncan}]. At a fixed point, $\b(\a)=0$, when we
set the physical electron mass equal to zero, the divergence of
the scale current is given by
 \be
\partial^{\m}D_{\m}= \frac{\b(\a)}{2\a}F_{\m\n}F^{\m\n} + [1+\g_{\theta}
(\a)\; ] \, m \, \bar{\psi} \psi = 0 \label{Eq2.6} \ee
 and hence scale
invariance is exact. Assuming that scale invariance is not
spontaneously broken, $Q_D|0>\, =\, 0$, where \be Q_D=\int\, d^3x
D_0({\bf x},t), \ee ({\it i.e.,\/} no dilatons are present in QED)
from which Eq.(\ref{Eq2.3}) follows.

Let us now consider the Dyson-Schwinger equation satisfied by the
inverse of the full, exact, renormalized electron propagator in
massless, PT symmetric QED:
 \be S^{-1}(p)= Z_2 \dirac p + iZ_2 e^2(2\pi)^{-4}\int
\; d^4k\; \g^{\m}D_{\m\n}S(k) \G^{\n}(p,k). \label{Eq2.8} \ee
 It
is well-known that at the fixed point $\beta(\alpha)=0$, in a
theory of $m=0$, spin-$\half$ QED, the full, exact, renormalized
``photon" propagator is  given exactly by the free ``photon"
propagtor, as established by Eguchi [\ref{Eguchi}] and thus we
have  in the Landau gauge: \be D_{\m\n}(q)=
\left(\frac{q_{\m}q_{\n}}{q^2} -g_{\m\n} \right )\frac{1}{q^2}\, ,
\ee \label{Eq2.9}where $q_{\m}=p_{\m}-k_{\m}$. This is so since
$\beta(\alpha)=\gamma_A(\alpha)=0$, where $\gamma_A (\alpha)= \mu
(\partial/ \partial \mu) \ln Z_3$ is the anomalous dimension of
the ``photon". The solution to the Ward-Takahashi identity
satisfied by the renormalized, proper vector vertex function in
QED can be determined in the standard manner by the gauge
technique [\ref{Delburgo}] to yield the form

\be \G^{\n}= \G^{\n}_L + \G^{\n}_T, \ee where the longitudinal
part of the vertex function admits the general, kinematical
singularity-free solution [ \ref{Balletal}, \ref{ourWTwork}] given
by \br \G^{\n}_L(p,k) &=& \half (A+\tilde{A}) \g^{\n} + \ds
{\frac{\S-\tilde{\S}}{k^2-p^2} } (\g^{\n}\dirac p + \dirac k
\g^{\n}
)\nonumber \\
&-& \half {\ds  \left (\frac{A - \tilde {A}  } {k^2-p^2} \right )
}\left [\; 2 \dirac p \g^{\n} \dirac k + (p^2 + k^2 ) \g^{\n} \;
\right ]\, ,  \label{Eq2.11} \er
 and the transverse piece obeys the
condition \be (p-p')_{\m}\G^{\m}_T(p,p')=0 \, , \ee and
consequently undetermined by the Ward-Takahashi identity. Here we
have employed the notation $\tilde{A}= A(k^2)$ etc. It must be
stressed that we shall retain the transverse piece throughout our
calculation. Indeed,  undetermined and arbitrary as it is, the
transverse piece of the vertex has a significant bearing on our
conclusions. In massless finite QED, when the chiral symmetry is
exact, the above solution for the vertex function reduces to
 \be
\G^{\n}(p,k) = \half (A+\tilde{A}) \g^{\n} - \half {\ds  \left (
\frac{A - \tilde {A}  } {k^2-p^2} \right ) } [\; 2 \dirac p
\g^{\n} \dirac k + (p^2 + k^2 ) \g^{\n} \; ] + \G^{\n}_T (p,k)\, ,
\ee and the Dyson-Schwinger equation, Eq.(\ref{Eq2.8}) reduces to
\be S^{-1}(p)=\dirac p A(p^2) = Z_2 \dirac p + iZ_2 e^2 (2
\pi)^{-4}\int\; d^4k \left (\frac{\dirac q q_{\n}}{q^2}- \g_{\n}
\right ) \frac{1}{q^2}\frac{\dirac k}{k^2 \tilde{A}} \left [
\G^{\n}_L(p,k) + \G_T^{\n}(p,k) \right ]. \label{Eq2.14}
\ee
 We shall now address the transverse part of the vertex function.
 The general invariant form of the transverse part has been
 determined by Ball and Chiu [\ref{Balletal}]. In the special case
 of massless electron, only four of the basic functions contribute
 [\ref{Bashir}] and accordingly, we have
 \begin{equation}\label{2.14a}
    \Gamma_T^{\nu}(p,k,q)=\sum_{i=2,3,6,8}\, \tau_i(p^2,k^2,q^2)\,
    T_i^{\nu}(p,k)\, ,
\end{equation}
where the  $\tau_i(p^2,k^2,q^2)$ are unknown, arbitrary functions
and the invariant functions are listed below:
\begin{eqnarray}
  T_2^{\nu} &=& (p^{\nu}k\cdot q -k^{\nu} p\cdot q)\, (\dirac k + \dirac p)
  \nonumber \,, \\
  T_3^{\nu} &=& q^2 \gamma^{\nu} -q^{\nu} \dirac q \nonumber\, , \\
  T_6^{\nu} &=& \gamma^{\nu}(k^2-p^2)- (k+p)^{\nu }(\dirac k - \dirac p)
  \nonumber\, , \\
  T_8^{\nu} &=& \gamma^{\nu} \sigma_{\lambda \nu}k^{\lambda}p^{\nu}
  -k^{\nu} \dirac p + p^{\nu}\dirac k \label{2.14b}\, .
\end{eqnarray}
We may now evaluate the trace over the Dirac matrices after
multiplying by $\dirac p$ and dividing by $4 p^2$. After a tedious
computation, we obtain the following result: \br &&A(p^2)- Z_2= -
iZ_2 e^2 (2\pi)^{-4}\int \; d^4k\;
\frac{1}{p^2k^2A(k^2)} \nonumber \\
& & \left \{ \frac{1}{2q^4}(A + \tilde{A}) [2 p^2k^2 - (p \cdot k)
(p^2+k^2)\;] + \frac{1}{q^4}\frac{A-\tilde{A}}{k^2-p^2}(p \cdot k)
(p^2-k^2)^2
\right. \nonumber \\
&&- \left. (A+\tilde{A}) \frac{(p \cdot k)}{q^2} - \frac{\left
(A-\tilde{A}\right )}{q^2(k^2-p^2)}\left [ 4 (p \cdot k)^2 - (p
\cdot k) (p^2+k^2) \right ]  + {\G}^1_T(p^2,k^2,p \cdot k)\right
\}, \label{Eq2.15} \er  where ${\G}^1_T$ arises from the trace
calculation of the term containing the transverse vertex piece and
is given by the expression
\begin{eqnarray}
  \frac{1}{q^4} \{~~&\tau_2&~p \cdot k (p^2-k^2) [k^2(p^2-p \cdot k)
  + 2 p \cdot k (p^2 - p\cdot k )-p^2 (p\cdot k - k^2)]
   \nonumber \\
     &&+ \tau_6 (p^2-k^2)[k^2(p^2-p\cdot k)-2 p\cdot k(p^2-p\cdot k)
   +p^2(p\cdot k - k^2)]  \nonumber\\
     &&- \tau_8 \half (k^2-p\cdot k)
   [2p\cdot k(p^2-p\cdot k)-p^2(p\cdot k -k^2)]
   -\tau_8\half k^2(p^2-p\cdot k)^2  \} \nonumber\\
   + \frac{1}{q^2}\{&& 2\tau_2p^2k^2
   (p\cdot k - k^2)(p \cdot k)[p^2(p\cdot k -k^2)
   -k^2(p^2 - p\cdot k)]+ 2 \tau_2 (p\cdot k)^2 (p\cdot k-k^2)  \nonumber\\
&&+3\tau_3 (p\cdot k)\nonumber\\
    &&+\tau_6 [p\cdot k (k^2-p^2) + 2\{  (p\cdot k)^2-p^2k^2\}
   +3 (p\cdot k)^2 k^2-p^2]
    \nonumber\\
   &&+3 \tau_8 [(p\cdot k)^2 - p^2k^2]  \}\label{2.15a}\, .\\
 \end{eqnarray}
This  non-linear integral equation satisfied by the function
$A(p^2)$  is the exact consequence of the Dyson-Schwinger equation
for the electron propagator in finite QED at the Gell-Mann-Low
fixed point since we have not introduced any approximations and we
have not discarded the transverse piece.  Transforming to
Euclidean momenta by implementing the transformations: \be d^4k
\rightarrow id^4 k = i\int\, k^3 dk d\Omega; \quad p^2 \rightarrow
-p^2; \quad k^2 \rightarrow -k^2; \quad p\cdot k \rightarrow -
p\cdot k. \ee
 and after performing the angular integrals,  we
obtain
 \br A(-p^2) - Z_2&=& Z_2 e^2
(2\pi)^{-4}\int \; k^3 dk\; \frac{1}{p^2k^2A(-k^2)}
\nonumber\\
& & \left \{\,  \half (A + \tilde{A})\left [ 2 p^2k^2 I_4 -
(p^2+k^2) I_5\; \right ] -
(p^2-k^2) (A-\tilde{A}) I_5 \right . \nonumber\\
&-& \left . (A+\tilde{A})I_2 - \frac{\left ( A-\tilde{A}\right )
}{(p^2-k^2)}\left [ 4 I_3 -  (p^2+k^2)I_2 \right ]  + {\tilde
\G}^1_T (-p^2,-k^2)\right \} \label{Eq2.17}\, , \er where
\begin{equation}\label{2.15a}
{{\tilde \G}}^1_T(-p^2,-k^2)= \int\, d\Omega \,
{\G}^1_T(-p^2,-k^2,p \cdot k)\, .
\end{equation}
 We
should recall that due to the dependence on $p\cdot k$ and the
fact that the invariant functions $\tau_i$ are arbitrary and
unknown, the angular integration of the transverse part cannot be
explicitly carried out in the \emph{exact} theory. The term
containing the transverse part, namely ${\tilde \G}^1_T
(-p^2,-k^2)$ is therefore an undetermined quantity. We have left
it in this form only to indicate that it contains nonzero terms
and henceforward we shall denote the consequences of this
transverse vertex part   appropriately as ${\tilde \G}^2_T,
{\tilde \G}^3_T $ etc. It must be stressed that the transverse
parts\emph{ must} survive:  their presence is essential for
multiplicative renormalizability of the propagator
[\ref{CornwallKing}]. The quantities $I_1, I_2 \cdots$ are angular
integrals listed in our earlier work, [\ref{RAPNS1}]. All momenta
are Euclidean, defined by $p_E=\sqrt{-p^2}, \; k_E=\sqrt{-k^2}$
and in what follows we shall drop the subscript $E$ for Euclidean
momenta in order to avoid clutter. Making use of the results in
Appendix A of our earlier work [
\ref{RAPNS1},\ref{ArnowittDeser}], we arrive at the following
result
 \br A(p^2)-Z_2 &=& - \frac{Z_2
e^2}{16\pi^2} \int_0^{\infty}\; k \, dk \frac{1}{p^2 A(k^2)} \\
\nonumber & & \left \{ \frac{A(p^2)-A(k^2)}{ (p^2-k^2)} \left [
\frac{2\s^2 (p^2-k^2)^2}{p^2_{>}(1-\s^2)}- p_{<}^2(1+\s^2) \right
] + {\tilde \G}^2_T(p^2, k^2) \right \}, \label{Eq2.18} \er
 where
${\tilde \G}^2_T(p^2, k^2)$ arises from the transverse vertex
part, $\s=p_{<}/p_{>}$  and $p_{<}= {\rm min} \{p,k\}, \, p_>=
{\rm max} \{p,k\}$.  This result is a consequence of the essential
ingredients of finite QED, with no approximations nor additional
assumptions. In order to ascertain whether  this finite theoy of
QED, as we have developed thus far, admits of a solution to the
Gell-Mann-Low eigenvalue equation, we proceed as follows. The
self-consistency of the theory constructed in this manner, of
massless QED at the fixed point can be checked by making the
replacement \be A(p^2)= \left ( \frac{p^2}{\m^2} \right )^{\g} \ee
in accordance with Eq.(\ref{Eq2.3}) where $\g=\g(\a)$ is the QED
anomalous dimension in the $m=0$ theory. After some algebra, we
thus obtain the result \br \left (  \frac{p^2}{\m^2} \right
)^{\g}-Z_2 &=& - \frac{Z_2 e^2}{16\pi^2} \int_{0}^{p}\; \,k\, dk
\left \{ [(p^2/k^2)^{\g}-1] \frac{(p^2k^2-3k^4)}
{p^4(p^2-k^2)} + {\tilde \G}_T^3(p^2,k^2) \right \} \nonumber \\
&-&  \frac{Z_2 e^2}{16\pi^2} \int_{p}^{\infty}\;   k \, dk \left
\{ [(p^2/k^2)^{\g}-1] \frac{(p^2k^2-3p^4)}{p^2k^2(p^2-k^2)}  +
{\tilde \G}_T^4(p^2,k^2) \right \},
 \er
where ${\tilde \G}_T^3$ and ${\tilde \G}_T^4$ represent the
contributions arising from the transverse piece vertex function.
With a change in variables, $s=p^2, k^2=sx$, this can be rewritten
in the form \br \left (  s/\m^2  \right )^{\g}-Z_2 &=& - \frac{Z_2
e^2}{32\pi^2} \int_{0}^{1}\; \left \{ \, dx \
\frac{x(1-3x)(x^{-\g}-1)} {(1-x)} +
\G_T^5(s,x) \right \} \nonumber \\
&-& \frac{Z_2 e^2}{32\pi^2} \int_1^{\infty}\; \left \{ dx\,
\frac{x(x-3)(x^{-\g}-1]}  {(1-x)}
 + \G_T^6(s,x) \right \},
\label{Eq2.21} \er where $\G_T^5$ and $\G_T^6$ (suppressing the
tilde henceforward) are contributions arising from the transverse
vertex piece. For general values of $\g$, we can evaluate the
integrals [\ref{RAPNS1}] and we obtain the result
 \br
 \left (  s/\m^2  \right )^{\g}&-&Z_2 = - \frac{Z_2
e^2}{32\pi^2 } \left \{
3F(1,3,4;1)  - F(1,2,3;1) + F(1,2-\g,3-\g;1) \right.\nonumber \\
 &-& 3F(1,3-\g,4-\g;1)+   3F(1,-1,0;1) -F(1,-2,-1;1)\nonumber \\
&+& \left. F(1,\g-2,\g-1;1) -3F(1,\g-1,\g;1)   + \G_T^7(s) \right
\} \label{Eq2.22} \er
 in terms of the hypergeometric functions,
where $\G^7_T(s)$ arises from the integral of the contribution
from the transverse vertex piece.

If we evaluate Eq.(\ref{Eq2.22}) at $s=\m^2$, we obtain \br
Z_2=1&+& \frac{Z_2 e^2}{32\pi^2 } \left \{
3F(1,3,4;1)  - F(1,2,3;1) + F(1,2-\g,3-\g;1) \right.\nonumber \\
 &-& 3F(1,3-\g,4-\g;1)+   3F(1,-1,0;1) -F(1,-2,-1;1)\nonumber \\
&+& \left. F(1,\g-2,\g-1;1) -3F(1,\g-1,\g;1)   + \G_T^7(\m^2)
\right \}, \label{Eq2.24} \er which can be rewritten as \br
Z_2^{-1}=1&-& \frac{e^2}{32\pi^2 } \left \{
3F(1,3,4;1)  - F(1,2,3;1) + F(1,2-\g,3-\g;1) \right.\nonumber \\
 &-& 3F(1,3-\g,4-\g;1)+   3F(1,-1,0;1) -F(1,-2,-1;1)\nonumber \\
&+& \left. F(1,\g-2,\g-1;1) -3F(1,\g-1,\g;1)   + \G_T^7(\m^2)
\right \}, \label{2.25} \er

This has been obtained in the Landau gauge to all orders in $\a$.
We are now ready to analyze the results contained in
Eqs.(\ref{Eq2.22}, \ref{2.25}), a major consequence of the
Dyson-Schwinger equations of finite QED with massless electron at
a fixed point, specifically in the PT symmetric QED. Care is
required in handling the  hypergeometric functions appearing in
these equations and the reader is referred to Appendix B of our
earlier work [\ref{RAPNS1}].

\vspace{.2in}
 \large
 \noindent \textbf{3. Conclusion and Summary}

\vspace{.2in} \normalsize

Let us examine Eq.(\ref{Eq2.22}) and determine what are the
allowed values of $\g$, the anomalous dimension in massless QED.
From Eqs.(\ref{Eq2.22}) and (\ref{Eq2.24}), we obtain \br \left (
s/\m^2  \right )^{\g}&=& 1 + \frac{Z_2 e^2}{32\pi^2 } \left \{
3F(1,3,4;1)  - F(1,2,3;1) + F(1,2-\g,3-\g;1) \right.\nonumber \\
 &-& 3F(1,3-\g,4-\g;1)+   3F(1,-1,0;1) -F(1,-2,-1;1)\nonumber \\
&+& \left. F(1,\g-2,\g-1;1) -3F(1,\g-1,\g;1)   + \G_T^7(\m^2) \right \}\nonumber \\
&-&  \frac{Z_2 e^2}{32\pi^2 } \left \{
3F(1,3,4;1)  - F(1,2,3;1) + F(1,2-\g,3-\g;1) \right.\nonumber \\
 &-& 3F(1,3-\g,4-\g;1)+   3F(1,-1,0;1) -F(1,-2,-1;1)\nonumber \\
&+& \left. F(1,\g-2,\g-1;1) -3F(1,\g-1,\g;1)   + \G_T^7(s) \right
\}, \label{3.1} \er
 which simplifies to
 \be
 \left (  s/\m^2
\right )^{\g}=  1 + \frac{Z_2 \alpha}{8 \pi} \left \{
\G_T^7(\m^2)- \G_T^7(s) \right \}. \label{3.2}
 \ee
We recall that the necessary and sufficient condition for finite
$Z_2$ and finite $Z_2^{-1}$ is $\gamma =0$. That is, if $\gamma
=0$, then both $Z_2$ and $Z_2^{-1}$ are finite. Conversely, if
$Z_2$ and $Z_2^{-1}$ are both finite, then $\gamma$ must vanish.
This assertion follows from the defining relation,
Eq.(\ref{Eq2.7}). In such a finite field theory, all the three
renormalization constants, $m_0, \, Z_2, \, $ and $Z_3$ tend to
finite limits in the limit of infinite cut-off parameter
$\Lambda$.  We can then observe that our stated result in Landau
gauge, with $\gamma =0$, yields
\begin{equation}\label{29}
    Z_2^{-1}(\mu^2, \xi =0) = 1 - \frac{\alpha}{8 \pi} \,
    \Gamma^7_T (\mu^2)\, .
\end{equation}
The divergences present in the longitudinal pieces clearly cancel
only if $\gamma =0$, for an \emph{arbitrary} choice of the
transverse piece $\Gamma^7_T (\mu^2)$. This has been demonstrated
in Appendix B of our earlier work [\ref{RAPNS1}]. From Eq.
(\ref{29}), we then conclude that $\Gamma^7_T (\mu^2)$ must also
be finite when $\gamma =0$. Furthermore, Eq.(\ref{3.2}) simplifies
to
\begin{equation}\label{30}
    Z_2(\xi=0)\,\frac{\alpha}{8 \pi}\left \{\Gamma^7_T (\mu^2) -
    \Gamma^7_T (s)\right \} =0
       \, .
\end{equation}
From Eqs.(\ref{29}) and (\ref{30}) we finally conclude that
\begin{equation}\label{31}
    e^2 \equiv 0\, ,
\end{equation}
since $Z_2(\xi=0)$ cannot vanish ($Z_2^{-1}$ must be finite) and
the transverse piece vertex \emph{cannot} be a constant
independent of $s$. In other words, the only unphysical manner
that one can arrive at a non-trivial eigenvalue $e^2\neq 0$ is if
the transverse vertex piece is a constant, but it is not.

We can provide an alternative demonstration of the trivial
eigenvalue result directly from Eq.(\ref{29}) as follows. From the
definition of $\gamma(\alpha)$, we thus obtain
\begin{equation}
\label{32} \gamma(\alpha)=-\mu \frac{\partial}{\partial \mu} \ln
\left ( 1 - \frac{\alpha} {8 \pi} \Gamma^7_T(\mu^2)\right )
 = \frac{1}{8\pi
(1- \frac{\alpha}{8 \pi}\Gamma^7_T (\mu^2))} \{ \beta \,(\alpha)\,
\Gamma^7_T(\mu^2) + \alpha \, \mu \frac{\partial
\Gamma^7_T(\mu^2)}{\partial \mu}\}\, .
\end{equation}
From Eq.(\ref{29}), $Z_2^{-1}$ is finite, $\beta (\alpha)= \mu
(\partial \alpha/ \partial \mu)=0$ at the fixed point, and it thus
follows that
\begin{equation}\label{33}
    \gamma (\alpha)= \frac{\alpha}{8 \pi}\, Z_2 \, \mu \frac{\partial \Gamma^7_T (\mu^2))}
    {\partial \mu}\, .
\end{equation}
Hence $\gamma (\alpha)=0$ implies $\alpha \equiv 0, \, \mu $
arbitrary and $\partial \Gamma^7_T(\mu^2)/\partial \mu \neq 0$.

We have carried out our investigation in the Landau gauge.  The
form of the solution $A(p^2)=(p^2/ \m^2)^{\g (\xi)}$ must remain
valid in all covariant gauges, $\xi\neq 0$, in the minimal
subtraction scheme [\ref{Zinnjustin}] at $\b(\a)=0$. It remains to
investigate the non-perturbative gauge technique in a general
setting and determine the choice of $\xi$ which will accomplish
the task of establishing a finite field theory non-perturbatively,
i.e., to all orders in $\alpha$. This task will occupy us in a
forthcoming work.

It may be important to emphasize that what we have examined is
non-perturbative QED and our conclusions are based on
non-perturbative techniques developed for investigating the
Dyson-Schwinger equation in QED, now extended to include the PT
symmetric field theory.

The PT symmetric QED is a theory with asymptotic freedom. We have
demonstrated that the trivial eigenvalue $\alpha =0$ is the only
solution. The beta function starts out negative, $\beta(\alpha) <
0$, at the origin, remains negative for all $\alpha > 0$ and does
not turn over to positive values since the beta function cannot
have an infrared fixed point. We therefore conclude that the PT
symmetric QED is a theory which incorporates both asymptotic
freedom and infrared slavery.

The central result that $\beta(\alpha) <0$ for all $\alpha >0$ may
be contrasted with the circumstance in non-supersymmetric Quantum
Chromodynamics (QCD) with a large number of flavors $N_F$, where a
non-trivial fixed point of $\beta(\alpha)$ does exist and the
theory is in a non-Abelian coulomb phase [\ref{BanksZaks}]. in
contrast, here in PT-QED, it is expected that infrared slavery
will lead to the generation of dynamical ``photon" mass, i.e., the
theory will exhibit a \emph{mass gap}. We shall address this issue
in a separate publication.

{\large {\bf References and Footnotes}}

\vspace{.2in} \normalsize
\begin{enumerate}
  \item \label{RAPNS1} R. Acharya and P. Narayana Swamy, Int. J.
Mod. Phys.\textbf{A 12} 3799 (1997).
 \item\label{JBW} K. Johnson and M.
Baker, Phys. Rev. {\bf D8} 1110 (1973) and references cited
therein;  M. Baker and K.Johnson, Physica {\bf A 96}, 120 (1979);
S.L.Adler, C.Callan, R.Jackiw and D.Gross, Phys.Rev. {\bf D6},
2982 (1972). See also N. Krasnikov, Phys.Lett. {\bf B225}, 284
(1989).
 \item \label{Adler} S.L.Adler, Phys. Rev.
{\bf D5}, 3021 (1972); J. Bernstein, Nucl.Phys.{\bf B95}, 461
(1975). See also S.L.Adler, preprint arXiv:hep-ph/0505177, May
2005.
 \item \label{Gellmannlow} M.
Gell-Mann and F. E. Low, Phys. Rev. {\bf 95}, 1300 (1954);
N.N.Bogoliubov and D.Shirkov, {\it Introduction to the theory of
quantized fields\/}, Interscience, New York, 1959; K. Wilson,
Phys. Rev. {\bf D3}, 1818 (1971).
\item \label{Adleranomaly} J.S.Bell and R.Jackiw, Nuovo Cimento
{\bf A60}, 47 (1969); S.L.Adler, Phys. Rev. {\bf 177}, 2426
(1969); R.Jackiw and K.Johnson, Phys. Rev. {\bf 182}, 1459 (1969);
S.L.Adler and W.Bardeen, Phys. Rev. {\bf 182}, 1517 (1969);
C.R.Hagen, Phys. Rev. {\bf 177}, 2622 (1969); B. Zumino, {\it
Proceedings of the Topical conference on Weak Interactions\/},
CERN, Geneva 1969), p.361.
 \item \label {BakerJohnson} M.Baker and K.Johnson, Phys.
Rev. {\bf D3}, 2516 (1971).
\item \label{Delburgo} A.Salam, Phys. Rev. {\bf 130}, 1287 (1963);
R.Delburgo and P. West, Phys. Lett. {\bf B72}, 3413; {\it ibid\/}
J.Phys. {\bf A 10}, 1049 (1977). See also D.W.Atkinson and
H.A.Slim, Nuovo Cimento {\bf A50}, 555 (1979).

 \item \label{Balletal} J.Ball and
F.Zachariasen, Phys. Lett.{\bf B 106}, 133 (1981); J.Ball and
T.Chiu, Phys. Rev. {\bf D 22}, 2542 (1980); J. Cornwall,
\emph{Phys. Rev.} \textbf{D 26}(1983), 1453.


\item \label{Bender} C.Bender et al, preprint
arXiv:hep-th/0501180, January 2005; C.Bender and S. Boettcher,
\emph{Phys. Rev. Lett.} \textbf{80}, 5243 (1998); C. Bender et al,
\emph{Phys. Rev. Lett.}\textbf{93}, 251601 (2004); C. Bender et
al, preprint arXiv:hep-th/0411064, Nov. 2004; other references
cited therein. See also M.E.Peskin and D.V.Schroeder, \emph{An
Introduction to Field theory}, p.425, Addison-Wesley Publishing
(1995) New York.

\item\label{WeinBook} S.Weinberg,\emph{The Quantum theory of
Fields}, Vol.II, p.154, Cambridge University Press (1996) New
York.

 \item\label{Witten} P. Argyres, M. Plesser, N. Seiberg and
E. Witten, \emph{Nucl. Phys.} \textbf{B 461},(1996) 71 .

\item \label{Bernstein} J.Bernstein, {\it Elementary Particles and
their Currents\/}, W.H.Freeman and Co., 1968.

 \item \label{ourNCpaper} R.Acharya and P.
Narayana Swamy, Nuovo Cimento {\bf A103}, 1131 (1990).

 \item \label{Weinberg} S. Weinberg, Phys.Rev. {\bf 118}, 838
(1960).

\item \label{AdlerBardeen} See {\it e.g.,\/} S.L.Adler and
W.Bardeen, Phys.Rev. {\bf D4},  3045 (1971).

 \item \label{Callan} C.
Callan, {\it Summer School of Theoretical Physics, Les Houche
1971\/} editors C.Dewitt and C.Itzykson. Gordon Breach publishers
(1973) New York; see also S. Weinberg, Phys.Rev. {\bf D8}, 3497
(1973).

\item \label{AdlerCollinsDuncan} S.Adler, J.C.Collins and
A.Duncan, Phys. Rev. {\bf D15}, 1712 (1977).

 \item \label{Eguchi}
T.Eguchi, Phys.Rev. {\bf D17}, 611 (1978).

\item \label{ourWTwork} It is more expedient to express the
solution in terms of the functions $A$ and $\S$: see {\it e.g.,\/}
R. Acharya and P. Narayana Swamy, Nuovo Cimento {\bf A98}, 773
(1987).

\item \label{Bashir} A.Bashir and R.Delbourgo, preprint
arXiv:hep-ph/0405018, May 2004.


 \item\label{CornwallKing}
J.M.Cornwall, \emph{Phys.Rev.} \textbf{D 26}, 1453 (1983);
J.E.King, \emph{Phys.Rev.}\textbf{D 27}, 1821 (1983).

 \item
\label{ArnowittDeser} Some of these results are contained in
R.Arnowitt and S.Deser, Phys. Rev. {\bf138B }, 712 (1965).


\item \label{Zinnjustin}J. Zinn-Justin, {\it Quantum Field theory
and critical phenomena\/}, second edition, Clarendon Press, (1993)
Oxford.
 \item\label{BanksZaks}
T.Banks and A.Zaks, \emph{Nucl.Phys} \textbf{B 196},189 (1982). It
has been estimated in QCD that $N_F^{\rm min}\approx 12$; see
T.Appelquist et al, \emph{Phys.Rev.} \textbf{D 58} , 105017
(1998).

\end{enumerate}

\end{document}